\documentclass[aps,pra,reprint]{revtex4-2}

\usepackage[english]{babel}
\usepackage{graphicx}
\usepackage{dcolumn}
\usepackage{bm}
\usepackage{braket}
\usepackage{xcolor}
\usepackage{float}
\usepackage{amsfonts}
\usepackage{dsfont}
\usepackage{verbatim}
\usepackage{amsmath}
\usepackage{cancel}
\usepackage[normalem]{ulem}

\usepackage{xr}
\usepackage{hyperref}

\usepackage[export]{adjustbox}  

\begin{document}

\preprint{APS/123-QED}

\title{Activating thermally charged quantum batteries in finite time: Thermodynamic trade-offs between correlations, work, and information}

\author{Bhaskar Barman}%
\affiliation{Department of Microtechnology and Nanoscience (MC2), Chalmers University of Technology, S-412 96 Göteborg, Sweden}
\affiliation{Indian Institute of Science Education and Research Kolkata, 741246 West Bengal, India}
\affiliation{Indian Institute of Technology Madras, 600036 Tamil Nadu, India}

\author{Janine Splettstoesser}
\affiliation{Department of Microtechnology and Nanoscience (MC2), Chalmers University of Technology, S-412 96 Göteborg, Sweden}

\author{Henning Kirchberg}
\email[]{henning.kirchberg@chalmers.se}
\affiliation{Department of Microtechnology and Nanoscience (MC2), Chalmers University of Technology, S-412 96 Göteborg, Sweden}

\date{\today}

\begin{abstract}
Thermally charged quantum batteries provide a stable and \textit{easy-to-have} energy resource, but remain passive and therefore do not allow useful energy extraction \textit{on demand} via unitary operations. We introduce a time‑dependent \textit{stirring} protocol that activates a thermally charged quantum battery through a time-dependent coupling to an auxiliary activator, thereby generating correlations that drive the battery into an active state. Accounting explicitly for the energetic cost of the stirring process, the entropic cost of correlation generation and the activation time, we derive bounds on the maximal net extractable energy. Incorporating projective measurement on the activator and exploiting the information gained through the measurement further enhances the extractable energy. As an experimentally relevant example, we analyze a waveguide-QED setup where a harmonic‑oscillator battery (waveguide) is  stirred and monitored by a two‑level system. We characterize the performance of the protocol in terms of net extractable energy (net ergotropy) and power output. 
\end{abstract}

\maketitle



\section{\label{intro}Introduction}
\begin{figure}[t]
    \centering
    \includegraphics[width=\linewidth]{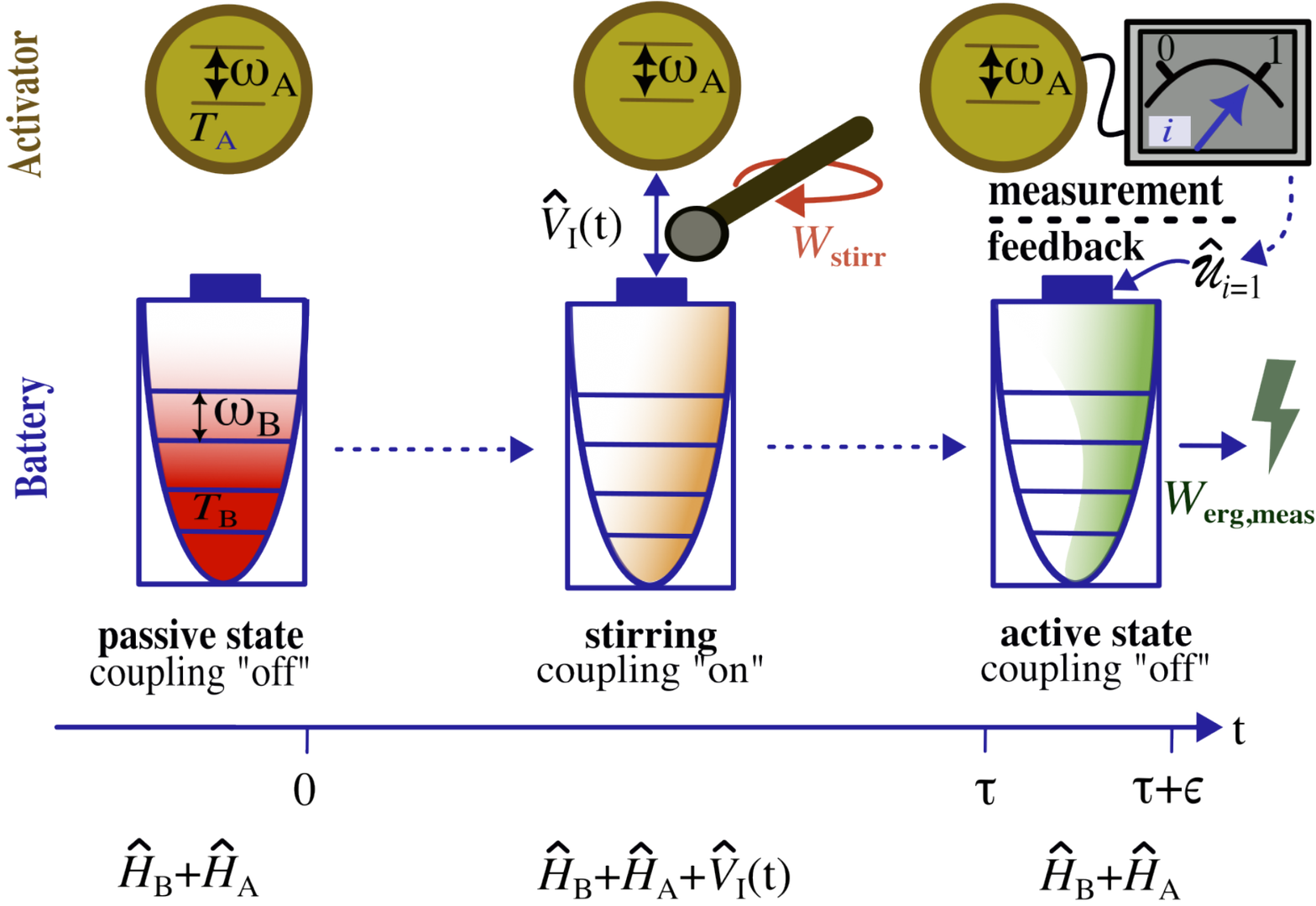} 
    
    \caption{Schematic of the activation process. An initially thermally-charged harmonic-oscillator \textit{quantum battery} (frequency $\omega_{\rm B}$ and Hamiltonian $\hat{H}_{\rm B}$), is prepared in a \textit{passive} thermal state at temperature $T_{\rm B}$. At time $t=0$, the battery is coupled to a two-level \textit{activator} (frequency $\omega_{\rm A}$ and Hamiltonian $\hat{H}_{\rm A}$) initialized in a thermal state at temperature $T_{\rm A}$, by the time-dependent \textit{stirring} Hamiltonian $\hat{V}_{\rm I}(t)$ and requiring a stirring work $W_{\rm stirr}$. At $t=\tau$, the battery and activator are decoupled. A projective measurement of the activator in the state $i=0$ or $i=1$ conditions a taylored, unitary energy-extraction operation $\hat{\mathcal{U}}_{i=0;1}$. The measurement, feedback, and energy extraction processes occur quasi-\textit{instantaneously}: their characteristic duration $\epsilon$ satisfies $\epsilon\ll2\pi/\omega_{\rm B} $ and $\epsilon\ll2\pi/\omega_{\rm A} $ and is also much faster than the thermalization with the respective thermal baths.}
    \label{fig:qb_setup}
\end{figure}
Quantum batteries are envisioned to store and release energy \textit{on demand} in quantum processes~\cite{Ferraro2026,Auffeves_2022,Campbell_2026,Kur_2026}, exploiting genuine quantum resources such as coherence and entanglement~\cite{Alicki2013,Binder_2015, Campaioli_2017}. 
Typically, the operational advantage of a quantum battery is limited by environmental disturbances. The decay of coherence and entanglement in the presence of thermal noise restricts the battery’s ability to store and deliver useful energy~\cite{NitzanBook,BreuerBook,Satriani_2024}. A natural strategy to address this fragility is to shift perspective: instead of treating the environment solely as a source of dissipation, one may regard ambient thermal energy as a resource that can be stored in the battery. In this spirit, the battery is allowed to equilibrate with its environment, effectively becoming thermally charged~\cite{Barra_2022,Feliu_2024,Satriani_2024,ahmadi2026}. Such states constitute a \textit{free resource}, as they are easy to prepare, stable, and require no external control~\cite{Chit_2019}.

This convenience, however, comes with a fundamental limitation. Standard dissipation-less energy-extraction protocols, such as those based on unitary control (e.g., stimulated emission)~\cite{Monsel_2020,Hagman_2025,Kirchberg_2025}, cannot extract work from thermal, i.e., \textit{passive}, states~\cite{allahverdyan_maximal_2004}. Consequently, a key challenge is to transform a thermally charged battery into an \textit{active} state prior to extraction, such that useful energy, quantified by the ergotropy, becomes accessible~\cite{allahverdyan_maximal_2004}.

A variety of mechanisms have been proposed to generate or sustain such active states. These include coherent driving via unitary operations~\cite{Campaioli_2017,Binder_2015}, coupling to auxiliary quantum systems~\cite{Ferraro_2018,Ferraro2026}, strong coupling to a single thermal reservoir~\cite{Hovhannisyan_2020}, and measurement-based protocols~\cite{Satriani_2024}. Another class of approaches exploits nonequilibrium steady states, maintained through chemical or temperature gradients, to induce activity in the battery~\cite{Tacchino_2020,Wang_2025}. 
Despite their diversity, all activation mechanisms require consumption of resources. For instance, maintaining external potentials demands sustained voltage supply, while coupling and decoupling to auxiliary systems entails non-negligible energetic costs. To remain practically useful, a battery must yield more energy than is required to generate or maintain its active state and to extract the stored energy~\cite{Monsel_2020}. This trade-off highlights the need for carefully designed activation strategies for thermally charged quantum batteries that minimize operational costs while enabling \textit{on-demand} energy extraction.

In this work, we build on these considerations and address a central challenge: how \textit{fast} can such a readily available \textit{thermal} energy resource be transformed into an active form that enables controlled useful energy extraction on demand by additionally minimizing the energetic and entropic costs? More generally, we seek to understand the finite-time thermodynamic trade-offs that govern the activation of thermally charged quantum batteries. We are motivated by the recent proposal to recycle residual energy from quantum computational processes, stored, for instance, in a waveguide mode~\cite{Kur_2026}. The quantum state associated with such stored energy is, in general, not expected to be active; rather, it is likely to be in thermal equilibrium with its environment arising from residual heating, for example due to dissipation through signal attenuation~\cite{Krinner2019}.

To this end, we propose an adiabatic work stroke, the \textit{stirring}, equivalent to that one in a quantum heat engine~\cite{Alicki_1979,Geva_1992,Gelbwaser_2013,deAssis2021} to activate the thermally charged quantum battery in finite time. The stirring occurs through a time-dependent unitary coupling of the battery to an auxiliary quantum system at initially different temperature, which we refer to as the activator, see Fig.~\ref{fig:qb_setup}. 
The stirring enables energy exchange between battery and activator. Depending on the direction of energy flow, the process admits interpretations analogous to an actual heat engine, a heat valve, or a refrigerator. After this stirring, the battery is in an active state where useful energy is extractable, see Fig.~\ref{fig:qb_setup}.

The stirring process generates correlations between the battery and the activator, which have been shown to be necessary for useful energy extraction via ergotropy from an otherwise passive battery state~\cite{Shi_2022}. These correlations can be further exploited by incorporating an additional measurement step. Projective measurements performed on the activator after the stirring provide information about the joint system state. This information can then be used through outcome-conditioned unitary operations to enhance the extracted energy~\cite{Francica2017}.

To identify an optimal activation strategy, it is therefore essential to analyze the interplay between the proposed fundamental resources involved in battery activation: (i) the entropic cost occurring due to the generation of battery–activator correlations, (ii) the energetic input required to couple and decouple the two systems; and (iii) the information acquired through measurements and its optimal conversion into additional extractable energy. These resources are intrinsically linked through the duration of the stirring stroke, making the activation time a key thermodynamic parameter. At the same time a heat flow is required that occurs since the battery and activator are initially prepared at different temperatures. 

In this work, we quantify the interrelation between these resources. By explicitly accounting for the associated energy costs and the activation time, we derive bounds on the maximal extractable net energy from the quantum battery in finite time. Our results thereby provide quantitative guidelines for on-demand energy extraction from residual energy originating in quantum processes.

This paper is organized as follows. In Sec.~\ref{sec:GeneralStirring}, we introduce a general stirring protocol between a quantum \textit{battery} and an \textit{activator} that transforms an initial thermal (passive) state into an active state from which useful energy can be extracted. We identify the thermodynamic operation regimes of the stirring process and discuss energy-extraction protocols based on general unitary operations (quantified by ergotropy). We provide a detailed analysis of the maximally extractable energy, both \textit{with} and \textit{without} measurement, accounting for the time-dependent energetic costs of the stirring process. In Sec.~\ref{sec:OscillatorQubit}, we present an experimentally relevant setup in which a harmonic-oscillator battery is stirred and monitored by a two-level activator. Also the \textit{power} output, namely the maximally extractable net energy per stirring time, is analyzed. 

\section{Stirring Protocol and resources}
\label{sec:GeneralStirring}

In this section, we describe the separate steps of the protocol for battery activation and energy extraction, based on a general model.

\subsection{Initialization}
We consider a quantum battery described by the Hamiltonian $\hat{H}_{\rm B}$. In the beginning of the proposed protocol, the battery is in contact with a thermal bath at temperature $T_{\rm B}$, see the left part of Fig.~\ref{fig:qb_setup} for $t<0$. The battery's initial state is hence given by the thermal (Gibbs) state
\begin{align}
\label{eq:sys_in}
\hat{\rho}_\mathrm{B}(0) = \frac{\exp(-{\hat H_{\rm B}}/{T_{\rm B}})}{{\rm{tr}}\{\exp(-{\hat H_{\rm B}}/{T_{\rm B}})\}},
\end{align}
where we have set the Boltzmann constant $k_\mathrm{B}\equiv1$.
The thermal state is \textit{passive}, meaning that no useful energy can be extracted from it by a unitary protocol. 

In order to prepare the battery for energy extraction, our protocol includes an active control unit, which we refer to as \textit{activator}. The activator is an ancillary quantum system described by a Hamiltonian $\hat{H}_{\rm A}$. It is initially coupled to its own thermal bath of temperature $T_{\rm A}$, which is in general different from the one of the bath connected to the battery. The activator's initial state is hence 
\begin{align}
\label{eq:act_in}
\hat{\rho}_\mathrm{A} (0) = \frac{\exp(-{\hat{H}_{\rm A}}/{T_{\rm A}})}{{\rm{tr}}\{\exp(-{\hat H_{\rm A}}/{T_{\rm A}})\}}.
\end{align}
 The total initial state is therefore captured by 
 \begin{equation}
     \hat \rho(0) = \hat \rho_{\rm B}(0)\otimes\hat\rho_{\rm A}(0)\ ,
 \end{equation}
since the battery and activator are initially not coupled to each other. 

\subsection{Stirring protocol}\label{sec:stirring}
At time $t=0$, the battery and activator are decoupled from their respective baths and coupled to each other during the time interval $0 < t < \tau$.
Even though in a practical situation, the battery and the activator cannot simply be ``decoupled" from their environment, this is a reasonable assumption as long as the time of the protocol is significantly shorter than the inverse system-bath couplings. In this limit, environmental effects can be neglected during the protocol, and the dynamics becomes effectively unitary. 

This protocol is analogous to the adiabatic work stroke of a quantum heat engine, during which no heat is exchanged with the environment and the system entropy remains constant~\cite{Alicki_1979}. The separation of timescales between the work stroke and relaxation to thermal equilibrium is well justified in qubit-waveguide platforms, where measurement and control operations can be performed much faster than the relevant relaxation and decoherence processes~\cite{Yang2025}. 

The adiabatic work stroke that couples and decouples the battery and activator is modeled by a time-dependent interaction Hamiltonian, $\hat{V}_{\rm I}(t)$, representing an externally controlled potential that is nonzero only for $0<t<\tau$. 
The battery-activator protocol is therefore expressed by the total Hamiltonian
\begin{equation}
\label{eq:model_general}
    \hat{H}(t) = \hat{H}_{\rm B} + \hat{H}_{\rm A} + \hat{V}_{\rm I}(t).
\end{equation}
The joint battery-activator dynamics during the work stroke in the absence of bath coupling is captured by the unitary evolution 
\begin{align}
\label{eq:density_time}
    \hat{\rho}(t) = \hat{U}(t)\hat{\rho}(0)\hat{U}^{\dagger}(t),
\end{align}
where $\hat{U}(t) = \mathcal{T} \exp[-\dfrac{i}{\hbar} \int_0^t dt' \hat{H}(t') ]$ and $\mathcal{T}$ is the time-ordering operator.
Importantly, since $[\hat{V}_{\rm I}(t), \hat{H}_{\rm B}+\hat{H}_{\rm A}] \neq 0$, energy is exchanged between the battery, the activator, and a work reservoir, which provides the externally tuned coupling $\hat V_{\rm I}(t)$. 
Assuming instant switching of the coupling Hamiltonian at times $t=0$ (``on") and at $t=\tau$ (``off"), the energy exchanged with the work reservoir is~\cite{Kirchberg_2025,Hagman_2025}
\begin{align}
\label{eq:stirr_cost}
W_{\rm stirr}(\tau) &\equiv {\rm{tr}}\{[\hat{H}_{\rm B} + \hat{H}_{\rm A}](\hat{\rho}(\tau) - \hat{\rho}(0))\} \\ \notag
& =:\Delta E_{\rm A}(\tau) +\Delta E_{\rm B}(\tau). 
\end{align}
Here, $\hat{\rho}(t)$ is given by Eq.~\eqref{eq:density_time} and the trace is taken over both subsystems, $\rm tr\equiv tr_Atr_B$. Note that the Hamiltonian entering the expression in Eq.~\eqref{eq:stirr_cost} corresponds to the \textit{total} Hamiltonian at times $t=0$ and $t=\tau$, where the battery-activator coupling is switched off, $\hat{V}_{\rm I}=0$.  

Depending on the sign of Eq.~\eqref{eq:stirr_cost}, work must either be supplied by the external work reservoir, $W_{\rm stirr}>0$, or can be extracted from the battery-activator system, $W_{\rm stirr}<0$. In the latter case, the amount of extracted work is determined by the specific choice of interaction protocol $\hat V_{\rm I}(t)$ and is generally not maximal for the given initial thermal states of the battery and activator. 

Rather than focusing on the non-optimal work extracted during the work stroke itself, we are interested in quantifying the \textit{maximal} useful energy extractable from the battery. 
In fact, the states of the battery and of the activator become correlated during the work stroke. 
We refer to this as \textit{stirring} of the battery by the activator, since it makes the battery state deviate from a thermal state. The result of this stirring is that the battery state $\hat{\rho}_{\rm B}(t)\equiv{\rm tr}_{\rm A}\hat{\rho}(t)$ is transformed into an active state from which maximally the reversible useful energy characterized via \textit{ergotropy}, see Sec.~\ref{subsec:WorkExtr}, can be extracted.

\subsection{Energy extraction and ergotropy}
\label{subsec:WorkExtr}

The purpose of the stirring process is to drive the battery away from thermal equilibrium and prepare it in an active state. Once the stirring stroke is completed at time $t=\tau$, we therefore ask how much useful energy can be \textit{maximally} reversibly extracted from the battery alone. 

We further assume that the duration of the extraction protocol, denoted by $\epsilon$, is much shorter than both the intrinsic dynamical timescales of the battery and activator, $2\pi/\omega_{\rm B}$ and $2\pi/\omega_{\rm A}$, respectively, as well as the characteristic bath-induced relaxation times. As illustrated in Fig.~\ref{fig:qb_setup}, the battery state instantaneously react to the extraction protocol.

The maximum useful energy that can be extracted in this manner is the \textit{ergotropy}~\cite{allahverdyan_maximal_2004}
\begin{align}
\label{eq:ergotropy_def}
W_{\rm erg} (\tau) \equiv \text{\rm tr}_{\rm B} \left( \hat \rho_{\rm B} (\tau) \hat H_{\rm B} \right) - \min_{\hat {\mathcal{U}}} \text{tr}_{\rm B} \left( \hat {\mathcal{U}} \hat \rho_{\rm B} (\tau) \hat{\mathcal{U}}^\dagger \hat H_{\rm B} \right).
\end{align}
The second term in this expression is minimized over all possible unitaries. 
Unitary processes are reversible and, by construction, do not generate entropy.
The so obtained optimal unitary $\hat{{\mathcal{U}}}_{\rm p}$ transforms the battery into a passive state $\hat {\mathcal{U}}_{\rm p} \hat \rho_{\rm B}(\tau) \hat {\mathcal{U}}^\dagger_{\rm p} \to \hat \rho_{{\rm B},p}(\tau)$. By construction we have $W_{\rm erg}\geq 0$ \cite{allahverdyan_maximal_2004}. In fact, if the battery remains in a passive state such as a thermal state, no energy can be extracted via a unitary operation and hence the ergotropy vanishes.

To highlight the role of the ergotropy as maximally extractable energy, we rewrite it in terms of non-equilibrium free-energy differences~\cite{biswas_extraction_2022},
\begin{align}
\label{eq:ergotropy_free_energy}
W_{\rm erg}(\tau) 
&= F(\hat \rho_{\rm B}(\tau)) - F (\hat \rho_{\rm B,p}(\tau)),
\end{align}
where we have defined the  non-equilibrium free energy $F(\hat \rho_{\rm B})={\rm tr}_{\rm B} \left( \hat \rho_{\rm B} \hat H_{\rm B}\right)-T_{\rm B}S(\hat \rho_{\rm B})$ with respect to the battery-bath temperature and we have exploited the invariance of the von-Neumann entropy, $S(\hat \rho_{\rm B})= - {\rm tr}_{\rm B}(\hat \rho_{\rm B}\ln{\hat \rho}_{\rm B})=S(\hat \rho_{\rm B,p}) $, under unitary transformation. 

Furthermore, by subtracting and adding the free energy of the initial thermal battery state $\hat \rho_{{\rm B}}(0)\equiv \hat \rho_{{\rm B,th}}$ in Eq.~\eqref{eq:ergotropy_free_energy} and exploiting the fact that the free energy of the thermal state is minimal, we find an upper bound on the (non-negative) ergotropy 
\begin{align}
\label{eq:ergotropy_bound}
W_{\rm erg} (\tau) &=F(\hat \rho_{\rm B}(\tau))-F(\hat \rho_{\rm B}(0)) \\ \notag &- [F (\hat \rho_{\rm B,p}(\tau))-F(\hat \rho_{\rm B}(0))] \\ \notag 
    &\leq \Delta E_{\rm B}(\tau) -T_\mathrm{B}\Delta S(\hat \rho_{{\rm B}}(\tau)) ,
\end{align}
which is the change of the nonequilibrium free energy of the battery caused by the stirring~\cite{footnote5}. In the last line of Eq.~\eqref{eq:ergotropy_bound}, we identify the change of von-Neumann entropy of the battery
\begin{align}
\label{eq:entropy_diff}
\Delta S(\hat \rho_{{\rm B}}(\tau)) &\equiv S(\hat \rho_{{\rm B}}(\tau)) -S(\hat \rho_{{\rm B}}(0)) \\ \notag &= \Delta_{\rm i}S(\hat \rho_{{\rm B}}(\tau)) + \Delta_{\rm r} S(\hat \rho_{{\rm B}}(\tau)),   
\end{align}
which can be further decomposed into its \textit{irreversible}, $\Delta_{\rm i}S(t)$, and \textit{reversible}, $\Delta_{\rm r} S(t)$, component~\cite{esposito_entropy_2010}, see Appendix~\ref{appendix:work_Extr}.

The irreversible component reads 
\begin{align}
\label{eq:entropy_irrev}
\Delta_{\rm i}S(\hat \rho_{{\rm B}}(\tau)) = D\Big[\hat \rho(\tau)||\hat \rho_{\rm B}(\tau)\otimes \hat \rho_{\rm A}(0)\Big]\geq 0, 
\end{align}
with $D[\hat \rho||\hat \rho^{'}]\equiv {\rm tr}\hat \rho \ln\hat \rho - {\rm tr}\hat \rho \ln \hat \rho'\geq0$ being the quantum relative entropy. Equation~\eqref{eq:entropy_irrev} represents the entropy production during the stirring protocol related to the build-up of correlation between the battery and the activator~\cite{esposito_entropy_2010,footnote4}, and is always non-negative. Furthermore, the reversible part of the battery's entropy change, identified with the energy change of the activator, see Appendix~\ref{appendix:work_Extr}, is given by
\begin{align}
\label{eq:entropy_rev}
\Delta_{\rm r} S(\hat \rho_{{\rm B}}(\tau))  &= {\rm tr}_{\rm A}((\hat \rho_{\rm A}(\tau) - \hat \rho_{\rm A}(0))\ln{\rho_{\rm A}(0)})\\ \notag &=\frac{1}{ T_{\rm A}} {\rm tr}_{\rm A}((\hat \rho_{\rm A}(0) - \hat \rho_{\rm A}(\tau))\hat H_{\rm A}) \\ \notag
& = -\frac{1}{ T_{\rm A}}\Delta E_{\rm A}(\tau) ,
\end{align}
which is the \textit{thermodynamic} entropy change of the activator due to its energy change. We have used $\hat \rho_{{\rm A}}(0)\equiv \hat \rho_{{\rm A,th}}$, see Eq.~\eqref{eq:act_in}. 
Substituting Eq.~\eqref{eq:entropy_diff} together with Eqs.~\eqref{eq:entropy_irrev} and \eqref{eq:entropy_rev} in the upper bound on the ergotropy~\eqref{eq:ergotropy_bound}, we find
\begin{align}
\label{eq:Ergotropy_bound_final}
W_{\rm erg}(\tau) \leq  T_{\rm B} \bigg\{ &-D\Big[\hat \rho(\tau)||\hat \rho_{\rm B}(\tau)\otimes \hat \rho_{\rm A}(0)\Big] \\ \notag &+\frac{1}{T_{\rm A}} \Delta E_{\rm A}(\tau) + \frac{1}{T_{\rm B}} \Delta E_{\rm B}(\tau) \bigg\}\equiv W_\mathrm{erg}^\mathrm{bound}.
\end{align}
This provides a bound on the ergotropy, namely on the maximum energy that can be made available, after the battery is stirred by the activator. The first term is the entropic penalty related to the creation of correlations between the battery and the activator. It has been shown that the creation of correlations is a \textit{necessary} condition to extract energy from a system initially in a thermal state~\cite{Shi_2022}; this first term in Eq.~\eqref{eq:Ergotropy_bound_final} is always negative, since the relative entropy is a non-negative quantity. 
This entropic penalty must be compensated for. In Eq.~\eqref{eq:Ergotropy_bound_final}, this compensation is provided by the remaining two terms, which quantify the temperature-weighted energy changes of the battery and the activator. In other words, overcoming the entropic penalty of the correlation requires suitable entropic contributions from these subsystems. This can be achieved by exploiting two distinct thermodynamic resources—namely, heat and work. To make this connection explicit, we rewrite
\begin{align}\label{eq:operation-regimes}
    &\frac{1}{T_{\rm A}} \Delta E_{\rm A}(\tau) + \frac{1}{T_{\rm B}} \Delta E_{\rm B}(\tau) = \nonumber\\
    & \frac{\Delta E}{2}\left(\frac{1}{T_{\rm A}}-\frac{1}{T_{\rm B}}\right)+\frac{W_{\rm stirr}}{2} \left(\frac{1}{T_{\rm A}}+\frac{1}{T_{\rm B}}\right)\ ,
\end{align}
with the exchanged energy $\Delta E=\Delta E_{\rm A}-\Delta E_{\rm B}$ between the activator and battery and the energy exchange with an external resource $W_{\rm stirr}=\Delta E_{\rm A}+\Delta E_{\rm B}$, above related to the work exchange during the stirring stroke~\eqref{eq:stirr_cost}. The first contribution is positive, when energy flows from hot to cold, the second contribution is positive when an external agent (here the time-dependent coupling) provides energy to the system. Thermodynamically, an optimal \textit{activator} hence behaves like a heat valve: a heat flow from hot to cold at the cost required for enabling the flow. Note that in principle, the ergotropy can be larger than zero even when the system during the activation process operates as a heat engine (heat flow from hot to cold while providing energy to the external agent) or as a refrigerator (using energy provided by the external agent to move energy from cold to hot). However, as Eq.~\eqref{eq:operation-regimes} shows, these regimes are less favorable when the goal is to activate a thermally charged battery for energy extraction \textit{after the activation process} (via the unitary $\hat {\mathcal{U}}$ given in Eq.~\eqref{eq:ergotropy_def} and not via the external drive of $\hat{V}_{\rm I}$ during the activation process), see also Appendix~\ref{app:operationPrinciples}.

\subsection{Net energy extraction}
\label{subsec:netExtr}

The stirring process converts the initially passive state of the battery into an active one, from which useful energy up to the amount given by the ergotropy can be extracted. To quantify the overall benefit, we balance the ergotropy with the work exchange during the stirring stroke and define the \textit{net extractable} energy as
\begin{equation}\label{eq:W_net}
    W_{\rm net}(\tau)\equiv W_{\rm erg}(\tau)-W_{\rm stirr}(\tau)\ .
\end{equation}
In the thermodynamically optimal heat-valve regime during the stirring stroke, the related stirring work is a required resource and always positive.
Using the bound~\eqref{eq:Ergotropy_bound_final} together with the expression of the stirring cost~\eqref{eq:stirr_cost}, we obtain an upper bound on the net extractable energy 
\begin{align}
\label{eq:network_bound}
    &W_{\rm net}^{\rm bound}(\tau)\equiv W_{\rm erg}^{\rm bound}(\tau)-W_{\rm stirr}(\tau)
    \\ \notag &=-T_{\rm B}D\Big[\hat \rho(\tau)||\hat \rho_{\rm B}(\tau)\otimes \hat \rho_{\rm A}(0)\Big] +\bigg[\frac{T_{\rm B}}{T_{\rm A}} -1 \bigg] \Delta E_{\rm A}(\tau) .
\end{align}
One immediately observes from this expression that in order to achieve a positive net extractable energy, $W_{\rm net}^\mathrm{bound}(\tau)\geq0$, the temperatures related to the initial battery and activator states must differ, i.e., $T_{\rm A}\neq T_{\rm B}$. This is expected since a non-equilibrium regime is required to prepare a system in a non-passive state.
To analyze the requirements on bath temperatures further, we rewrite the condition of positive net energy extraction, $W_{\rm net}^\mathrm{bound}(\tau)\geq0$, as 
\begin{align}
\label{eq:bound_term}
    \bigg[\frac{T_{\rm B}}{{T_{\rm A}}} -1 \bigg] \Delta E_{\rm A}(\tau)> T_{\rm B}D\big[\hat \rho(\tau)||\hat \rho_{\rm B}(\tau)\otimes \hat \rho_{\rm A}(0)\big]\geq 0.
\end{align}
This  shows that if $\Delta E_{\rm A}<0$, we require $T_{\rm A}>T_{\rm B}$ to be able to extract net energy. Then, the left-hand side of~\eqref{eq:bound_term} is bounded by $|\Delta E_{\rm A}|$ for $T_{\rm A}\gg T_{\rm B}$.

Conversely, $\Delta E_{\rm A} > 0$ requires $T_{\rm A} < T_{\rm B}$. In this regime, the left-hand side of~\eqref{eq:bound_term} diverges as $T_{\rm A} \ll T_{\rm B}$, indicating that an energy flow into a cold activator, which hence acts as an entropy sink, favors net energy extraction under the proposed stirring protocol, confirming the insight from Eq.~\eqref{eq:operation-regimes}. Even though the battery loses energy, the process leverages the thermodynamically favorable heat flow from the \textit{hotter} battery (entropy source) to the \textit{colder} activator, rendering part of the battery’s internal energy accessible for energy extraction via ergotropy. Using Eqs.~\eqref{eq:ergotropy_bound},~\eqref{eq:W_net} and~\eqref{eq:stirr_cost}, one then sees that the net ergotropy is at most
\begin{equation}
      W_{\rm net}(\tau)\leq   -T_\mathrm{B}\Delta S(\hat \rho_{{\rm B}}(\tau)) - \Delta E_{\rm A}(\tau)
\end{equation}
namely the energy made extractable via the battery's entropy decrease minus the energy dumped into the activator.

\subsection{Measurement as resource}\label{sec:measurement}

The stirring protocol exploits correlations between the battery and activator which are created due to the external interaction potential $\hat V_{\rm I}(t)$. At the same time, this ``stirring" comes with a work cost that reduces the net extractable energy.  In order to leverage the created correlations for net useful energy extraction, we here analyze the opportunities of measurement and feedback. 

To design \textit{feedback} for extracting useful energy, see Fig.~\ref{fig:qb_setup}, we perform a \textit{projective measurement} on the state of the activator instantaneously after the stirring process.
To be concrete, at time $\tau$ the coupling between battery and activator is turned off and the state of the activator is determined by an instantaneous projective measurement of its state $\ket{i}_{\rm A}$ defined by the measurement operator $\hat\Pi_i\equiv \mathbb{I}_{\rm B}\otimes \ket{i}\bra{i}_{\rm A}$.
We then exploit the measurement result to choose a subsequent unitary protocol $\hat{\mathcal{U}}_i$ that is designed to best fit the given measurement result $i$ to extract energy. The maximum useful energy that can be extracted in this way was introduced as daemonic ergotropy in~\cite{Francica2017DaemonicCorrelations}.
It is given by
\begin{align}
\label{eq:dm_erg}
W_{\rm erg,meas}(\tau) =  \sum_{i} P(i,\tau) &\Big[ {\rm{tr}}\{\hat{\rho}(\tau|i)\hat{H}_{\rm B}\} \\ \notag & - \min_{\hat{\mathcal{U}}_i}{\rm{tr}} \{\hat{\mathcal{U}}_i \hat{\rho}(\tau|i)\hat{\mathcal{U}}^\dagger_{i}\hat{H}_{\rm B} \}\Big].    
\end{align}
with the density matrix conditioned on the measurement outcome, which reads
\begin{align}\label{eq:cond_rho_g}
\hat\rho(\tau|i) = \frac{\hat\Pi_i \hat \rho(\tau)\hat\Pi_i^\dagger}{\rm{tr}\{ \hat\Pi_i \hat \rho(\tau) \}}=\frac{\hat\Pi_i \hat \rho(\tau)\hat\Pi_i^\dagger}{P(i,\tau)}.
\end{align}
Following the same methodology as for the ergotropy, see Appendix~\ref{appendix:work_Extr}, we find an equivalent bound on the ergotropy in the presence of a measurement by 
\begin{align}
\label{eq:Dem_Ergotropy_bound_final}
W_{\rm erg,meas}(\tau) &\leq \rm  T_{\rm B} \bigg\{ -D\Big[\hat \rho(\tau)||\hat \rho_{\rm B}(\tau)\otimes \hat \rho_{\rm A}(0)\Big] \\ \notag &+I(\tau)+\frac{\Delta E_\mathrm{A}(\tau)}{T_{\rm A}} + \frac{\Delta E_\mathrm{B}(\tau)}{T_{\rm B}} \bigg\}\equiv W_{\rm erg,meas}^\mathrm{bound}.  
\end{align}
Importantly, this bound involves the non-negative information gain resulting from the measurement, see Appendix~\ref{appendix:work_Extr}
\begin{align}
\label{eq:InfoGain}
    I(\tau)= \sum_i P(i,\tau) {\rm tr}\big[ \hat{\rho}(\tau|i)\{\ln{\hat{\rho}(\tau|i)}-\ln{\hat{\rho}_{\rm B}(\tau)}\}\big]\geq0.
\end{align}
As expected, the information gain due to measurement can act as additional resource to overcome the cost creating the correlations between battery and activator, the latter being captured by the relative entropy in Eq.~\eqref{eq:Dem_Ergotropy_bound_final}. 

We find the \textit{net useful energy} extraction including the information obtained by measurement in analogy to Eq.~\eqref{eq:W_net}
\begin{equation}\label{eq:W_meas_net}
    W_{\rm net,meas}(\tau)\equiv W_{\rm erg,meas}(\tau)-W_{\rm stirr}(\tau)
\end{equation}
to be bounded by
\begin{align}
\label{eq:network_bound_dem}
    W_{\rm net,meas}^{\rm bound}(\tau)&\equiv W_{\rm erg,meas}^{\rm bound}(\tau)-W_{\rm stirr}(\tau)
    \\ \notag &=-T_{\rm B}D\Big[\hat \rho(\tau)||\hat \rho_{\rm B}(\tau)\otimes \hat \rho_{\rm A}(0)\Big] \\ \notag &+\bigg[\frac{T_{\rm B}}{T_{\rm A}} -1 \bigg] \Delta E_{\rm A}(\tau) +T_{\rm B}I(\tau) .
\end{align}
In the following, we analyze the stirring and measurement protocol at a concrete, experimentally relevant example.

\section{Energy-Harnessing from a qubit-activated oscillator}
\label{sec:OscillatorQubit}

We now demonstrate the charging and energy-extraction scheme introduced in Sec.~\ref{sec:GeneralStirring} at the experimentally relevant example of a battery realized by a harmonic oscillator and an activator realized by a two-level system. Indeed, such a setup can experimentally be realized, e.g., in  QED setups where energy can temporally be stored in a waveguide---typically captured by a harmonic mode---that could be stirred by a qubit~\cite{Linpeng2024,Zhang_2025,Dass2026} to activate the battery before energy extraction. We limit our discussion to the thermodynamically favorable heat valve regime, see Sec.~\ref{subsec:WorkExtr}, and choose our parameters accordingly. Other operation regimes are shown and discussed in Appendix~\ref{app:operationPrinciples}.

\subsection{Oscillator and qubit model}

The battery (harmonic oscillator) is hence described by the Hamiltonian 
\begin{align}
    \hat{H}_{\rm B}=\hbar\omega_{\rm B}\hat{b}^{\dagger}\hat{b},
\end{align}
 with frequency $\omega_\mathrm{B}$, and raising and lowering operators $\hat{b}^\dagger$ and $\hat{b}$, which follow standard bosonic commutation relations. Furthermore, the activator (qubit) is modeled by 
\begin{align}
    \hat{H}_{\rm A}=\hbar\omega_{\rm A}\hat{\sigma}_{\rm z}/2,
\end{align}
with frequency $\omega_{\rm A}$ and $\hat{\sigma}_{\rm z}$ being the Pauli $z$-matrix. In the initial state, the harmonic oscillator and the qubit are decoupled from each other and they are both in thermal equilibrium with their respective baths according to Eqs.~\eqref{eq:sys_in} and \eqref{eq:act_in}. At time $t=0$ the proposed stirring protocol starts, such that the composite system becomes correlated. We choose their time-dependent coupling to be of experimentally feasible Jaynes-Cummings form, see e.g.~\cite{Zhang_2025}, 
\begin{align}
\label{eq:JamesCummingsCoup}
    \hat{V}_{\rm I}(t)= \lambda(t) [\hat{b}\otimes\hat{\sigma}^+ +\hat{b}^{\dagger}\otimes \hat{\sigma}^{-}],
\end{align}
where $\hat{\sigma}^+$ and $\hat{\sigma}^-$ are the raising  and lowering operators of the qubit. The time-dependent coupling is instantaneously switched on at time $t=0$ to a constant value $\lambda$ and immediately switched off at time $t=\tau$, namely
\begin{equation}
    \lambda(t)=\hbar \lambda [\Theta(t)-\Theta(t-\tau)].
\end{equation}
This form of coupling is motivated by qubit–waveguide architectures where the waveguide hosts a selected photonic mode that couples to the qubit, with the interaction strength being dynamically controllable through an external magnetic flux~\cite{Janzen2023}.

\subsection{Dependence of extractable net energy on stirring time}
\begin{figure}[t!!]
\begin{center}
    \includegraphics[width=\columnwidth]{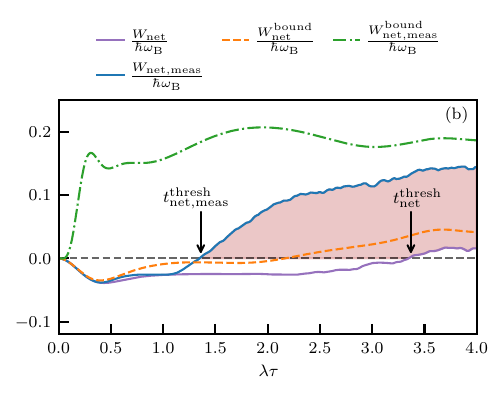}\\
  \includegraphics[width=0.97\columnwidth]{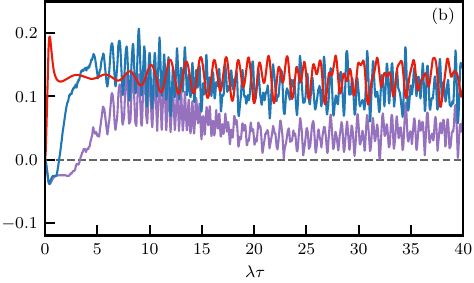}    
\end{center}
\caption{(a)~Net extracted useful energy with measurement, $W_{\rm net,meas}$, and without measurement, $W_{\rm net}$, plotted against stirring time $\lambda \tau$. Also their respective bounds, $W_{\rm net,meas}^{\rm bound}$, and, $W_{\rm net,meas}^{\rm bound}$, are shown. $W_{\rm net}$ and $W_{\rm net,meas}$ become positive at the threshold times, $t^{\rm{thresh}}_{\rm net}$ and $t^{\rm{thresh}}_{\rm net,meas}$, respectively. (b)~$W_{\rm net}$, $W_{\rm net,meas}$ and internal energy change $\Delta E_{\rm A}$ of the activator over a longer stirring time. Battery and activator parameters are set to $\hbar\omega_\mathrm{A}/T_\mathrm{A}=1.2$,  $T_{\rm A}/T_{\rm B}= 0.15$, $\omega_{\rm A}/\omega_{\rm B}= 1.2$, $\lambda/\omega_{\rm B}=0.1$, and $W_{\rm stirr}>0$ (identified as heat valve regime~\eqref{eq:conditionA} during the stirring stroke). 
}
    \label{fig:W_ext_vs_lambda_t}
\end{figure}

\begin{figure*}[t]
    \centering
    \includegraphics[height=7.9cm,width=0.49\linewidth]{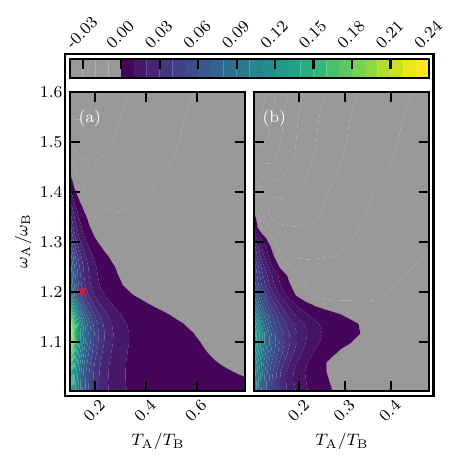}%
    \hfill
    \includegraphics[height=7.9cm,width=0.49\linewidth]{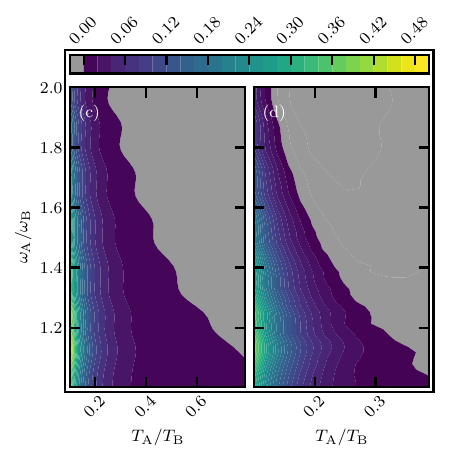}%
    
    \caption{
    (a) $W_{\rm net}^{\rm bound}$ ,
(b) $W_{\rm net}$ ,
(c) $W_{\rm net,meas}^{\rm bound}$ and 
(d)  $W_{\rm net,meas}$
as functions of relative frequency 
$\omega_{\rm A}/\omega_{\rm B}$ 
and relative temperature $T_{\rm A}/T_{\rm B}$.  We use $\tau = 4.0\,\lambda^{-1}$, $\hbar \omega_\mathrm{B}/T_\mathrm{B}=0.1$, and $\lambda/\omega_{\rm B}=0.01$. The red cross in (a) indicates the values for $\omega_{\rm A}/\omega_{\rm B}$  and  $T_{\rm A}/T_{\rm B}$ chosen in Fig.~\ref{fig:W_ext_vs_lambda_t}.
}
    \label{fig:allplots}
\end{figure*}

To exemplify how the proposed stirring protocol allows net energy extraction from a battery, which is initially in a thermal state, we calculate the net extractable energy -- both without measurement, see Eq.~\eqref{eq:W_net}, and with measurement, see Eq.~\eqref{eq:W_meas_net}. We start by analyzing how the extractable energy depends on the stirring time, namely on the time during which battery and activator are coupled to each other. The results are shown in Fig.~\ref{fig:W_ext_vs_lambda_t}.

We observe that the net extractable useful energy becomes positive after operation times of the order of the inverse coupling strength $\lambda$ between battery and activator. This time is indicated by the threshold times $t_\mathrm{net}^\mathrm{thresh}$ and $t_\mathrm{net,meas}^\mathrm{thresh}$, respectively. After this time the cost invested in the stirring is overcome by the useful energy extractable by ergotropy. When the ``stirring" is complemented by measurement, followed by feedback on the unitary that is chosen for the energy extraction, the threshold time is shortened, $t_\mathrm{net,meas}^\mathrm{thresh}<t_\mathrm{net}^\mathrm{thresh}$.  In addition, the amount of net extractable useful energy is significantly increased, when measurement and feedback is used. Indeed, we find $W_{\rm net,meas}\geq W_{\rm net}$ for any finite operation times, i.e., as expected information obtained by measuring the activator is an additional resource. Also the bounds on the net extractable energy without, Eq.~\eqref{eq:network_bound}, and with measurement, Eq.~\eqref{eq:network_bound_dem}, are shown. 

For the chosen set of parameters,  $\hbar \omega_\mathrm{B}/T_\mathrm{B}=0.1$, and the temperature of the battery bath being higher than the temperature of the activator, ${T_{\rm A}/T_{\rm B}= 0.1}$, we observe that after an initial increase, both $W_{\rm net}$ and $W_{\rm net,meas}$ exhibit fluctuations around a positive mean value as the stirring time increases. To understand the fluctuating behavior of $W_{\rm net}$ and $W_{\rm net,meas}$ with longer stirring time, it is useful to examine the change in the activator’s internal energy $\Delta E_{\rm A}$, as it enters the bound on $W_{\rm net}^{\rm bound}$, Eq.~\eqref{eq:network_bound}, and  $W_{\rm net,meas}^{\rm bound}$, Eq.~\eqref{eq:network_bound_dem}.  The initial relative occupation of the battery directly enters in the change of the activator's internal energy $\Delta E_{\rm A}$.

$\Delta E_{\rm A}$ is given by (see Appendix~\ref{app:internalEnergy} for details) 
\begin{subequations}
\begin{align}
\label{eq:internalEnergy}
    \frac{\Delta E_{\rm A}(\tau)}{\hbar\omega_\mathrm{A}}& =  \sum_m \frac{(2\lambda)^2(m+1)\left(p^\mathrm{A}_0p^\mathrm{B}_{m+1}-p^\mathrm{A}_1p^\mathrm{B}_m\right)}{\left(\omega_{\rm A}-\omega_{\rm B}\right)^2 + (2\lambda)^2(m+1)} \\ &\notag \times \sin^2{\left(\frac{\sqrt{\left(\omega_{\rm A}-\omega_{\rm B}\right)^2 + (2\lambda)^2(m+1)}}{2}\tau\right)},
\end{align}
where the initial thermal populations of the battery eigenstates $m$ and of the activator states are
\begin{eqnarray}
\label{eq:populationBattery}
    p^\mathrm{B}_m & = & \frac{e^{-\hbar\omega_{\rm B}m/T_{\rm B}}}{\sum_m e^{-\hbar\omega_{\rm B}m/T_{\rm B}}}\\
\label{eq:populationActivator}
    p^\mathrm{A}_1 & = & \frac{e^{-\hbar\omega_{\rm A}/T_{\rm A}}}{1+e^{-\hbar\omega_{\rm A}/T_{\rm A}}}
\end{eqnarray}
\end{subequations}
with $p^\mathrm{A}_0=1-p^\mathrm{A}_1$. Its temporal evolution is hence a sum of sinusoidal terms, oscillating in time, with weights depending on the occupation probabilities of the activator and the oscillator. 
Figure~\ref{fig:W_ext_vs_lambda_t}(b) shows the evolution of $\Delta E_{\rm A}(\tau)$ as a function of the stirring time. At short times, the dynamics are dominated by contributions from energy levels with large $m$, see Eq.~\eqref{eq:internalEnergy}, whereas at longer times progressively more levels with lower $m$ become relevant. How many energy eigenstates can actually contribute depends on the battery temperature through $p_m^\mathrm{B}$, which increases with $T_{\rm B}$ when keeping $\omega_{\rm B}$ constant. This results in oscillatory contributions that partially cancel or accumulate, producing fluctuations around an approximately constant positive value at long times, Fig.~\eqref{fig:W_ext_vs_lambda_t}(b).

\subsection{Conditions for positive net energy extraction}

We now study the condition for positive net energy extraction for different relative temperatures $T_{\rm A}/T_{\rm B}$ and the ratio of energy level splittings $\omega_{\rm A}/\omega_{\rm B}$ of the battery and activator for a fixed stirring time, $\tau=4/\lambda$.

We begin by analyzing the net energy extraction from the battery without measurement and feedback. The bound, $W_{\rm net}^{\rm bound}$, Eq.~\eqref{eq:network_bound} and $W_{\rm net}$, Eq.~\eqref{eq:W_net} are shown in Figs.~\ref{fig:allplots}(a) and~\ref{fig:allplots}(b), respectively.

For the specific qubit-waveguide setup, the \textit{general} necessary condition $\Delta E_\mathrm{A}[T_\mathrm{B}/T_\mathrm{A}-1]>0$ for \textit{positive net} useful energy extraction identified in Eq.~\eqref{eq:bound_term}, translates into 
\begin{subequations}
    \begin{eqnarray}
   \Delta E_{\rm A}(\tau)>0: & \ &   T_{\rm A}<T_{\rm B} \quad \mathrm{and} \quad \frac{\omega_{\rm A}}{\omega_{\rm B}}>1\label{eq:condition1},\\
      \Delta E_{\rm A}(\tau)<0: & \ &   T_{\rm A}>T_{\rm B} \quad \mathrm{and} \quad \frac{\omega_{\rm A}}{\omega_{\rm B}}<1\label{eq:condition2},
\end{eqnarray}
\end{subequations}
see Appendix~\ref{app:internalEnergy} for details. As discussed previously, see Sec.~\ref{subsec:WorkExtr}, an activator that is initially colder than the battery, condition~\eqref{eq:condition1}, allows for higher net energy extraction, while the battery is losing part of its initial thermal energy $\Delta E_{\rm B}<0$ in the correlation process due to stirring stroke from the outside applied potential. The actually achievable net energy extraction, $W_{\rm net}$, Fig.~\ref{fig:allplots}(b), follows the same qualitative behavior as $W_{\rm net}^{\rm bound}$, Fig.~\ref{fig:allplots}(a). As expected, the maximal net energy extraction is achieved when the activator temperature is small compared to that of the battery and their energy levels are comparable, i.e., $\omega_\mathrm{A} \approx \omega_\mathrm{B}$. Namely, in the special case $\omega_\mathrm{A} = \omega_\mathrm{B}$, the energy changes exactly compensate, $\Delta E_{\rm A} = -\Delta E_{\rm B}$ (see Appendix~\ref{app:internalEnergy}) and $W_{\rm stirr}$, Eq.~\eqref{eq:stirr_cost}, vanishes. In this regime, the transition of the battery into an active state is purely driven by the temperature difference and the related heat flow $\Delta E$, Eq.~\eqref{eq:operation-regimes}, between the \textit{hot} battery and \textit{cold} activator.

We next consider the net extractable energy from the battery when the activator is measured after the stirring process and the measurement result is used for the choice of the optimal unitary for energy extraction, see Sec.~\ref{sec:measurement}. In Figs.~\ref{fig:allplots}(c) and \ref{fig:allplots}(d), we plot $W_{\rm net,meas}$ and its corresponding bound $W_{\rm net,meas}^{\rm bound}$. Both quantities reach absolute values that are approximately twice as large as in the absence of measurement, highlighting the role of measurement and the related information gain as an additional resource to the battery–activator stirring process. 

\begin{figure}[t!!]
    \centering
    \includegraphics[width=0.95\linewidth]{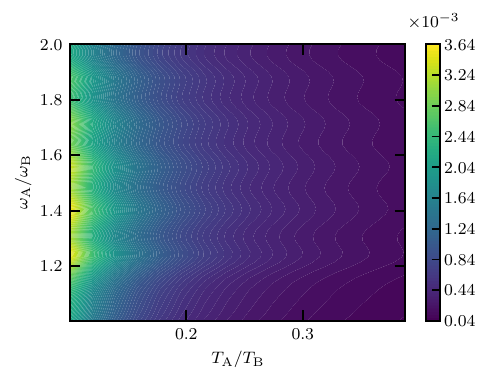}
    \caption{Information gain $I(\tau)$, Eq.~\eqref{eq:InfoGain}, plotted against $\omega_{\rm A}/\omega_{\rm B}$ and $T_{\rm A}/T_{\rm B}$ for $\tau = 4.0\lambda^{-1}$, $\hbar \omega_{\rm B}/T_{\rm B} = 0.1$, and $\lambda/\omega_{\rm B}=0.01$.}
    \label{fig:info_gain}
\end{figure}

\begin{figure}[t!!!]
    \centering
    \includegraphics[width=0.8\linewidth]{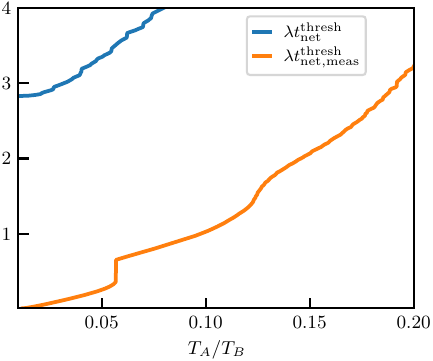}
    \caption{$\lambda\,t_{\rm net}^{\rm thresh}$ and $\lambda\,t_{\rm net, meas}^{\rm thresh}$ plotted against the relative temperature $T_{\rm A}/T_{\rm B}$ at $\omega_{\rm A}/\omega_{\rm B} = 1.4$, $\hbar \omega_\mathrm{B}/T_\mathrm{B}=0.07$, and $\lambda/\omega_{\rm B}=0.1$, and $W_{\rm stirr}>0$ (identified as heat valve regime~\eqref{eq:conditionA}). It shows the advantage of net energy extraction time in the presence of a measurement over the net energy extraction time.}
    \label{fig:thresold_time}
\end{figure}

In particular, $\omega_{\rm A}/T_{\rm A} > 1$ implies that the activator is cold and initially very likely in its ground state. Consequently, the dynamics of the joint battery-activator during stirring generates enhanced correlations, leading to an increased gain of information $I(\tau)$, Eq.~\eqref{eq:InfoGain}, as also seen in Fig.~\ref{fig:info_gain}. This, in turn, directly enhances the bound on the extractable useful energy given in Eq.~\eqref{eq:network_bound_dem}. Remarkably, despite this cost, $W_{\rm stirr} > 0$ , for correlation generation, net energy extraction---defined as the energy extracted from the battery minus this energetic investment, Eq.~\eqref{eq:W_meas_net}---is maximized in this operation regime. 

Finally, we analyze the conditions on the stirring time for positive energy extraction, quantified by the threshold times $t_{\rm net}^{\rm thresh}$ and $t_{\rm net,meas}^{\rm thresh}$ at which $W_{\rm net}^{\rm bound}$ and $W_{\rm net,meas}^{\rm bound}$, respectively, become positive. As shown in Fig.~\ref{fig:thresold_time}, both threshold times decrease with decreasing $T_{\rm A}/T_{\rm B}$. This highlights again that a cold activator relative to the battery acts as an entropic (cold) sink, facilitating useful energy extraction. The threshold time is further reduced in the presence of measurement and information gain, highlighting its role as an additional resource. We note that the threshold times exhibit slight non-monotonic features, as visible in Fig.~\ref{fig:thresold_time} particularly for $t_{\rm net,meas}^{\rm thresh}$. These arise from the oscillatory time dependence of $W_{\rm net}^{\rm bound}$ and $W_{\rm net,meas}^{\rm bound}$, which becomes more pronounced with increasing $T_{\rm B}$ (or, equivalently, decreasing $\hbar\omega_{\rm B}/T_{\rm B}$).

\subsection{Power from battery}
\label{subsec:PowerBattery}
To investigate the role of the stirring duration, we consider the net extractable useful energy generated per unit stirring time. We therefore define the power as
\begin{align}
    \label{eq:Power}
    P(\tau):=\frac{W_{\rm net,meas}(\tau)}{\tau}.
\end{align}
Here, the stirring time $\tau$ is assumed to be the dominant timescale of the protocol, while both the energy-extraction stage and the subsequent rethermalization occur on significantly shorter timescales. This assumption is justified for finite-size batteries and activators and has been discussed in detail for a related measurement-assisted energy-extraction protocol in Ref.~\cite{Hagman_2025,Kirchberg_2025}. In the following, we restrict our analysis to the measurement-assisted net extractable energy.

\begin{figure}[t!]
    \centering
    \includegraphics[width=\columnwidth]{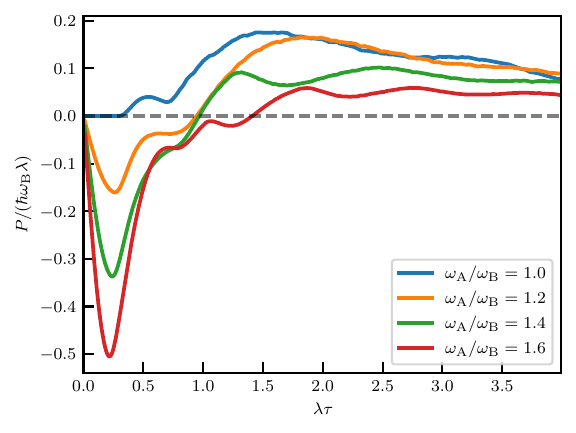}
    \caption{
        Net power output $P$ by choosing the relative activator to battery temperature $T_{\rm A}/T_{\rm B}=0.1$, $\hbar \omega_\mathrm{B}/T_\mathrm{B}=0.1$, $\lambda/T_{\rm B}=0.01$, and $\Delta E_{\rm A}(\tau)>0$ (for the chosen set of parameters).}
    \label{fig:Net_extr_power_vs_protocol_time_fixed_rel_freq}
\end{figure}

Fig.~\ref{fig:Net_extr_power_vs_protocol_time_fixed_rel_freq} shows the resulting power output $P$ in the regime $T_{\rm A}<T_{\rm B}$, $\omega_{\rm A}/\omega_{\rm B}>T_{\rm A}/T_{\rm B}$ and $\Delta E_A(\tau)>0$, Eq.~\eqref{eq:condition1}. As expected, a finite stirring time is required to obtain a nonzero power output, since the battery must first be driven into an active state through its coupling to the activator.

For short stirring times, the power is negative ($P < 0$), because the work cost of stirring, $W_{\rm stirr}$, exceeds the extractable energy from the battery. The stirring work increases with the ratio $\omega_{\rm A}/\omega_{\rm B}$ and vanishes only when $\omega_{\rm A} = \omega_{\rm B}$, in which case $P \geq 0$. A larger ratio $\omega_{\rm A}/\omega_{\rm B}$ therefore requires a higher external stirring cost to drive the battery into an \textit{active} state. As the stirring time $\tau$ increases, the battery is pushed further into this \textit{active} state, and the ergotropy eventually overcomes the stirring cost, resulting in a finite positive power output.

\section{Conclusion and Outlook}
\label{sec:conclusion}

We have demonstrated that a thermally charged quantum battery, although passive and therefore incapable of providing \textit{on-demand} energy through unitary operations alone, can be activated by a controlled time-dependent stirring stroke that couples it to an auxiliary activator.

We first identified battery--activator correlations as a necessary resource for activating the battery. The generation of correlations takes time and inevitably carries an entropic cost, and we showed that this cost is compensated most efficiently when heat flows from the hotter battery to the colder activator while additional work is supplied by the external control field. This operational regime, which we termed the \emph{heat-valve regime}, provides the most favorable conditions for battery activation.

A second essential resource is the work exchanged with the external potential that implements the time-dependent stirring protocol. By explicitly accounting for this energetic cost, we derived a general upper bound on the maximum net extractable useful energy. This bound is determined by the entropy reduction of the battery and the energy transferred to the activator, thereby linking thermodynamic irreversibility directly to the performance of the activated battery.

We further showed that the correlations generated during stirring can be harnessed as an informational resource. Projective measurements on the activator provide information about the battery state, enabling the application of conditional unitary operations that unlock additional extractable useful energy in the form of ergotropy. As a result, positive net useful energy can be achieved at significantly shorter stirring times, while both the net energy output and the charging power are substantially enhanced.

Finally, we analyzed an experimentally relevant implementation consisting of a harmonic-oscillator battery stirred by a qubit activator. This model reveals accessible operating regimes, particularly for a hot battery coupled to a colder activator, where positive net useful energy and finite power can be obtained after a finite activation time. Such conditions are well within reach of current waveguide-QED platforms.

Our results establish fundamental operational limits and design principles for activating thermal quantum batteries. More broadly, they highlight how useful energy can emerge from the interplay of energy exchange, correlation generation, information gain, and protocol duration, providing a route toward the \textit{on-demand} extraction of useful energy from thermally charged quantum devices.

\section*{Acknowledgment}
We thank Sofia Sevitz and Ludovico Tesser for carefully reading our manuscript and providing valuable feedback.
Funding from the European Research Council (ERC) under the Horizon Europe research and innovation program of the European Union (101088169/NanoRecycle) is gratefully acknowledged. 
\section*{Appendix}
\appendix

\section{Useful energy extraction and resources}
\label{appendix:work_Extr}

The ergotropy bound in Eq.~\eqref{eq:ergotropy_bound} is given by the change in the nonequilibrium free energy of the battery~\cite{footnote5}. It can be rewritten as
\begin{align}
\label{eq:app_erg_bound}
W_{\rm erg}(\tau)
&\leq -T_{\rm B}\Delta S(\tau)+\Delta E_{\rm B}(\tau)
\\ \notag
&= -T_{\rm B}
D\!\left[\hat\rho(\tau)\middle\|\hat\rho_{\rm B}(\tau)\otimes\hat\rho_{\rm A}(0)\right]
\\ \notag
&\quad
+{\rm tr}\!\left[
\left(\frac{T_{\rm B}}{T_{\rm A}}\hat H_{\rm A}\right)
\bigl(\hat\rho(\tau)-\hat\rho(0)\bigr)
\right] \\ \notag &
\quad+\Delta E_{\rm B}(\tau)
\\ \notag
&= -T_{\rm B}
D\!\left[\hat\rho(\tau)\middle\|\hat\rho_{\rm B}(\tau)\otimes\hat\rho_{\rm A}(0)\right]
+\frac{T_{\rm B}}{T_{\rm A}}\Delta E_{\rm A}(\tau) \\ \notag &
\quad+\Delta E_{\rm B}(\tau)
\\ \notag
&=
-T_{\rm B}\Delta_{\rm i}S(\tau)
-T_{\rm B}\Delta_{\rm r}S(\tau)
+\Delta E_{\rm B}(\tau).
\end{align}

Here,
\begin{equation}
\Delta_{\rm r}S(\tau)
=
\frac{1}{T_{\rm A}}
\,{\rm tr}_{\rm A}
\!\left[
\bigl(\hat\rho_{\rm A}(0)-\hat\rho_{\rm A}(\tau)\bigr)\hat H_{\rm A}
\right]
\end{equation}
denotes the reversible contribution to the battery entropy change associated with energy flow from the activator~\cite{esposito_entropy_2010}. Furthermore,
\begin{equation}
\Delta_{\rm i}S(\tau)
=
D\!\left[
\hat\rho(\tau)
\middle\|
\hat\rho_{\rm B}(\tau)\otimes\hat\rho_{\rm A}(0)
\right]
\end{equation}
is the irreversible entropy production~\cite{esposito_entropy_2010}, which originates from the correlations established between the battery and the activator. The quantum relative entropy is defined as
\begin{equation}
D[\hat\rho\|\hat\rho']
=
{\rm tr}(\hat\rho\ln\hat\rho)
-
{\rm tr}(\hat\rho\ln\hat\rho')
\ge 0.
\end{equation}

The bound on the ergotropy in the presence of measurement, Eq.~\eqref{eq:Dem_Ergotropy_bound_final}, can be derived in an analogous manner
\begin{align}
\label{eq:ergotropy_bound2}
W_{\rm erg,meas}(\tau)
&\leq
\Delta E_{\rm B}(\tau)
+
T_{\rm B}
\sum_i P(i,t)
\Bigl\{
{\rm tr}\!\left[\hat\rho(\tau|i)\ln\hat\rho(\tau|i)\right]
\\ \notag
&
-
{\rm tr}\!\left[\hat\rho_{\rm B}(0)\ln\hat\rho_{\rm B}(0)\right]
\Bigr\}
\\ \notag
&=
\Delta E_{\rm B}(\tau)
+
\frac{T_{\rm B}}{T_{\rm A}}
\Delta E_{\rm A}(\tau)
\\ \notag
&
-
T_{\rm B}
D\!\left[
\hat\rho(\tau)
\middle\|
\hat\rho_{\rm B}(\tau)\otimes\hat\rho_{\rm A}(0)
\right]
\\ \notag
&
+
T_{\rm B}
\sum_i P(i,t)\,
{\rm tr}
\!\left[
\hat\rho(\tau|i)
\bigl(
\ln\hat\rho(\tau|i)
-
\ln\hat\rho_{\rm B}(\tau)
\bigr)
\right]
\\ \notag
&=
\Delta E_{\rm B}(\tau)
+
\frac{T_{\rm B}}{T_{\rm A}}
\Delta E_{\rm A}(\tau)
\\ \notag
&
-
T_{\rm B}
D\!\left[
\hat\rho(\tau)
\middle\|
\hat\rho_{\rm B}(\tau)\otimes\hat\rho_{\rm A}(0)
\right]
+
T_{\rm B}I(\tau).
\end{align}

The additional contribution
\begin{equation}
I(\tau)
:=
\sum_i P(i,t)\,
{\rm tr}
\!\left[
\hat\rho(\tau|i)
\bigl(
\ln\hat\rho(\tau|i)
-
\ln\hat\rho_{\rm B}(\tau)
\bigr)
\right]
\ge 0
\end{equation}
quantifies the information gain obtained from the projective measurement of the activator outcome $i$. Since it is a weighted sum of relative entropies, $I(\tau)$ is manifestly non-negative.

\section{Operation principles during stirring (activation) stroke}
\label{app:operationPrinciples}

During the stirring stroke, the battery and the activator are coupled while initially prepared in thermal states at temperatures $T_{\rm B}$ and $T_{\rm A}$, respectively. The resulting temperature bias, enabled by the externally driven interaction protocol $\hat V_{\rm I}(t)$, gives rise to energy exchange between the two subsystems. The external modulation is associated with the stirring work $W_{\rm stirr}$ defined in Eq.~\eqref{eq:stirr_cost}, while the corresponding energy changes of the activator and battery are denoted by $\Delta E_{\rm A}$ and $\Delta E_{\rm B}$.

Although our primary objective is to maximize the useful energy that can subsequently be extracted from the battery, quantified by the ergotropy $W_{\rm erg}$, Eq.~\eqref{eq:ergotropy_def}, rather than the work exchanged during the activation process itself, it is nevertheless instructive to analyze the activation stroke from the perspective of thermodynamic operating principles~\cite{Manzano_2020,Monsel_2020}. Such an analysis allows us to classify the energy-transfer mechanisms occurring during battery activation.

Based on the direction of the energy flows and the sign of the stirring work, we identify three regimes of the activation process
\begin{align}
    \notag
    \mathrm{(a)\ heat\ valve}:\   
   &|\Delta E_{\rm A(B)}|>|\Delta E_{\rm B(A)}|;\ \Delta E_{\rm B(A)}<0;\\ 
   &\Delta E_{\rm A(B)}>0;\ T_{\rm B(A)}>T_{\rm A(B)};\ W_{\rm stirr}>0, \label{eq:conditionA}\\
   \notag \mathrm{(b)\ heat\ engine}:\ & 
  |\Delta E_{\rm A(B)}|<|\Delta E_{\rm B(A)}|;\ \Delta E_{\rm B(A)}<0;\\ 
   &\Delta E_{\rm A(B)}>0;\ T_{\rm B(A)}>T_{\rm A(B)};\ W_{\rm stirr}<0, \label{eq:conditionB} \\
   \notag
         \mathrm{(c)\ refrigerator}:\ &  
   |\Delta E_{\rm B(A)}|>|\Delta E_{\rm A(B)}|;\ \Delta E_{\rm B(A)}<0;\\ 
   &\Delta E_{\rm A(B)}>0;\ T_{\rm B(A)}<T_{\rm A(B)};\ W_{\rm stirr}>0 \label{eq:conditionC}.
\end{align}

\begin{figure}[h!!!]
    \centering
    \includegraphics[height=5.5cm, valign=t]{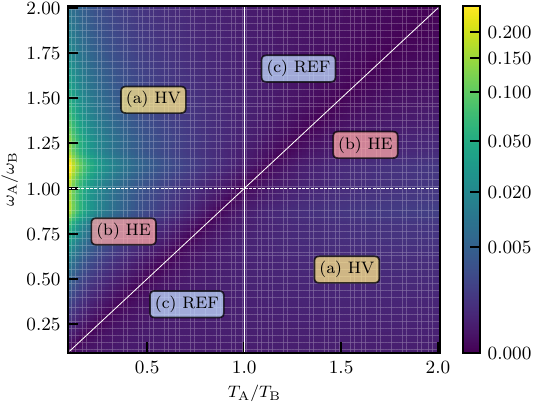}%

    \caption{
 $W_{\rm erg}^{\rm bound}$ 
as functions of relative frequency 
$\omega_{\rm A}/\omega_{\rm B}$ 
and relative temperature $T_{\rm A}/T_{\rm B}$. The parameters are chosen to be $\tau = 4.0\,\lambda^{-1}$, $\hbar \omega_\mathrm{B}/T_\mathrm{B}=0.1$, and $\lambda/\omega_{\rm B}=0.01$.
}
    \label{fig:allplot}
\end{figure}
We now investigate these operational regimes for the model discussed in Sec.~\ref{sec:OscillatorQubit}, consisting of a harmonic-oscillator battery coupled to a qubit activator. Notably, all three regimes allow for the activation of extractable useful energy in the battery.

Figure~\ref{fig:allplot} shows the ergotropy bound $W_{\rm erg}^{\rm bound}$ as a function of the relative temperature $T_{\rm A}/T_{\rm B}$ and the ratio of level spacings $\omega_{\rm A}/\omega_{\rm B}$. The parameters are chosen as $\tau = 4.0\,\lambda^{-1}$, $\hbar\omega_{\rm B}/T_{\rm B}=0.1$, and $\lambda/\omega_{\rm B}=0.01$. The figure also identifies the three thermodynamic operating regimes, which for the oscillator-qubit system can be characterized by the conditions derived in Appendix~\ref{app:internalEnergy}:

\begin{align}
    \mathrm{(a)\ heat\ valve}:\   
   & T_{\rm B(A)}>T_{\rm A(B)};\ \omega_{\rm A(B)}>\omega_{\rm B(A)};\\ \notag & \frac{\omega_{\rm A(B)}}{\omega_{\rm B(A)}}>\frac{T_{\rm A(B)}}{T_{\rm B(A)}}, \label{eq:conditionA1}\\
\mathrm{(b)\ heat\ engine}:\ & T_{\rm B(A)}>T_{\rm A(B)};\ \omega_{\rm A(B)}<\omega_{\rm B(A)};\\ \notag &  \frac{\omega_{\rm A(B)}}{\omega_{\rm B(A)}}>\frac{T_{\rm A(B)}}{T_{\rm B(A)}}, \label{eq:conditionB1} \\
         \mathrm{(c)\ refrigerator}:\   
   & T_{\rm B(A)}<T_{\rm A(B)};\ \omega_{\rm A(B)}>\omega_{\rm B(A)}\\ \notag &  \frac{\omega_{\rm A(B)}}{\omega_{\rm B(A)}}>\frac{T_{\rm A(B)}}{T_{\rm B(A)}} \label{eq:conditionC1}.
\end{align}
As follows from the ergotropy bound in Eq.~\eqref{eq:ergotropy_bound}, the heat-valve regime is particularly favorable for activating the battery. In this regime, the energy transfer induced during the activation stroke maximizes the upper bound on the extractable useful energy and therefore provides the most advantageous operating conditions for subsequent energy extraction.
\begin{widetext}
\section{Change in internal energy of battery and activator}
\label{app:internalEnergy}
For the harmonic-oscillator battery and qubit activator discussed in Sec.~\ref{sec:OscillatorQubit}, the changes in the internal energies of both subsystems during the stirring stroke can be obtained analytically. During this stroke, the battery and activator interact according to the unitary evolution

\begin{equation}
\hat{\rho}(\tau)
=
\hat U(\tau)\hat{\rho}(0)\hat U^\dagger(\tau),
\end{equation}

with the initially uncorrelated state

\begin{equation}
\hat{\rho}(0)
=
\hat{\rho}_{\rm B}(0)\otimes\hat{\rho}_{\rm A}(0),
\end{equation}
of the initial thermal battery and activator, Eqs.~\eqref{eq:sys_in} and~\eqref{eq:act_in}.

The unitary propagator generated by the Jaynes-Cummings interaction Hamiltonian in Eq.~\eqref{eq:JamesCummingsCoup} takes the form

\begin{equation}
\begin{aligned}
\hat U(\tau)
=&\,
|e\rangle\langle e|
\otimes
\Bigg[
\cos\!\left(
\frac{\sqrt{(\omega_{\rm A}-\omega_{\rm B})^2+(2\lambda)^2(\hat n+1)}}
{2}\tau
\right)
\\
&\qquad
-i\,
\frac{\omega_{\rm A}-\omega_{\rm B}}
{\sqrt{(\omega_{\rm A}-\omega_{\rm B})^2+(2\lambda)^2(\hat n+1)}}
\sin\!\left(
\frac{\sqrt{(\omega_{\rm A}-\omega_{\rm B})^2+(2\lambda)^2(\hat n+1)}}
{2}\tau
\right)
\Bigg]
\\
&+
|g\rangle\langle g|
\otimes
\Bigg[
\cos\!\left(
\frac{\sqrt{(\omega_{\rm A}-\omega_{\rm B})^2+(2\lambda)^2\hat n}}
{2}\tau
\right)
\\
&\qquad
+i\,
\frac{\omega_{\rm A}-\omega_{\rm B}}
{\sqrt{(\omega_{\rm A}-\omega_{\rm B})^2+(2\lambda)^2\hat n}}
\sin\!\left(
\frac{\sqrt{(\omega_{\rm A}-\omega_{\rm B})^2+(2\lambda)^2\hat n}}
{2}\tau
\right)
\Bigg]
\\
&-
i\,2\lambda\,\hat\sigma_+\hat b\,
\frac{
\sin\!\left(
\frac{\sqrt{(\omega_{\rm A}-\omega_{\rm B})^2+(2\lambda)^2\hat n}}
{2}\tau
\right)}
{\sqrt{(\omega_{\rm A}-\omega_{\rm B})^2+(2\lambda)^2\hat n}} - i\,2\lambda\,\hat\sigma_-\hat b^\dagger\,
\frac{
\sin\!\left(
\frac{\sqrt{(\omega_{\rm A}-\omega_{\rm B})^2+(2\lambda)^2(\hat n+1)}}
{2}\tau
\right)}
{\sqrt{(\omega_{\rm A}-\omega_{\rm B})^2+(2\lambda)^2(\hat n+1)}} ,
\end{aligned}
\end{equation}
where $\hat{\sigma}^+$; $\hat{b}^\dagger$ and $\hat{\sigma}^-$; $\hat{b}$ are the raising  and lowering operators of the qubit and oscillator, respectively. $\hat n=\hat b^\dagger \hat b$ denotes the oscillator number operator while $|e\rangle\langle e|=\hat{\sigma}^+\hat{\sigma}^-$ and $|g\rangle\langle g|=\hat{\sigma}^-\hat{\sigma}^+$. Using this propagator, the changes in the internal energies of the battery and activator can be evaluated analytically, yielding the results presented below.

In particular, the energy change of the battery is given by

\begin{align}
\Delta E_{\rm B} (\tau)   &= {\rm tr}\{\hat H_{\rm B}(\rho(\tau) - \rho(0))\} \label{eq:intEnergyBatt}
\\ \notag
& = \omega_{\rm B} p^\mathrm{A}_1 \sum_m \Bigg[mp_m^\mathrm{B} + \frac{(2\lambda)^2 (m+1)p_m^\mathrm{B}}{\left(\omega_{\rm A}-\omega_{\rm B}\right)^2 + (2\lambda)^2(m+1)}\sin^2{\bigg(\frac{\sqrt{\left(\omega_{\rm A}-\omega_{\rm B}\right)^2 + (2\lambda)^2(m+1)}}{2}\tau\bigg)} \Bigg]\\ \notag 
&+ \omega_{\rm B} p^\mathrm{A}_0 \sum_m \Bigg[mp_m^\mathrm{B} - \frac{(2\lambda)^2 m p_m^\mathrm{B}}{\left(\omega_{\rm A}-\omega_{\rm B}\right)^2 + (2\lambda)^2 m}\times \sin^2{\bigg(\frac{\sqrt{\left(\omega_{\rm A}-\omega_{\rm B}\right)^2 + (2\lambda)^2 m}}{2}\tau \bigg)} \Bigg]-\omega_{\rm B} \sum_m mp_m^\mathrm{B}
\\ \notag
&  =\omega_{\rm B} \sum_m \frac{(2\lambda)^2(m+1)}{\left(\omega_{\rm A}-\omega_{\rm B}\right)^2 + (2\lambda)^2(m+1)}(p^\mathrm{A}_1p_m^\mathrm{B} - p^\mathrm{A}_0p_{m+1}^\mathrm{B}) \sin^2\bigg({\frac{\sqrt{\left(\omega_{\rm A}-\omega_{\rm B}\right)^2 + (2\lambda)^2(m+1)}}{2}\tau }\bigg),
\end{align}
with the thermal population $p^{\rm B}_m$ of the battery states $m$ and activator $p_0^{\rm A}=1-p_1^{\rm A}$ defined in Eqs.~\eqref{eq:populationBattery} and~\eqref{eq:populationActivator}, respectively. 
\end{widetext}
Since all factors multiplying
$
p^\mathrm{A}_1p_m^\mathrm{B}
-
p^\mathrm{A}_0p_{m+1}^\mathrm{B}
$
are non-negative, a sufficient and necessary condition for the battery to gain energy, $\Delta E_{\rm B}(\tau)\geq0$, is that
\begin{align}
p^\mathrm{A}_1 p_m^\mathrm{B}
\geq
p^\mathrm{A}_0 p_{m+1}^\mathrm{B}.
\end{align}
 Substituting the thermal populations of the qubit activator~\eqref{eq:populationActivator} and harmonic-oscillator battery~\eqref{eq:populationBattery} yields

\begin{align}
\frac{
\left(
e^{-\hbar\omega_{\rm A}/T_{\rm A}}
-
e^{-\hbar\omega_{\rm B}/T_{\rm B}}
\right)
\left(
1-e^{-\hbar\omega_{\rm B}/T_{\rm B}}
\right)}
{1+e^{-\hbar\omega_{\rm A}/T_{\rm A}}}
\geq 0.
\end{align}

Because
$
1-e^{-\hbar\omega_{\rm B}/T_{\rm B}}>0
$
and
$
1+e^{-\hbar\omega_{\rm A}/T_{\rm A}}>0
$,
the condition simplifies to

\begin{align}
e^{-\hbar\omega_{\rm A}/T_{\rm A}}
\geq
e^{-\hbar\omega_{\rm B}/T_{\rm B}},
\end{align}

or, equivalently,

\begin{align}
\frac{T_{\rm A}}{T_{\rm B}}
\geq
\frac{\omega_{\rm A}}{\omega_{\rm B}}.
\label{eq:battery_charging_condition}
\end{align}

Equation~\eqref{eq:battery_charging_condition} therefore determines the parameter regime in which energy flows into the battery during the stirring stroke. 

The corresponding change in the internal energy of the activator is
\begin{widetext}
\begin{align}
\Delta E_{\rm A}  &= {\rm tr}\{\hat H_{\rm A}(\rho(\tau) - \rho(0))\} \label{eq:InternalEnergyActivator} \\ \notag 
& = \frac{\omega_{\rm A}}{2}\sum_{m=0}^{\infty} (p^\mathrm{A}_0p_{m+1}^\mathrm{B} -p^\mathrm{A}_1) \frac{(2\lambda)^2 (m+1)}{\left(\omega_{\rm A}-\omega_{\rm B}\right)^2 + (2\lambda)^2(m+1)} \sin^2{\bigg(\frac{\sqrt{\left(\omega_{\rm A}-\omega_{\rm B}\right)^2 + (2\lambda)^2(m+1)}}{2}\tau\bigg)} \\ \notag 
& -\frac{\omega_{\rm A}}{2}\sum_{m=0}^{\infty} (p^\mathrm{A}_1-p^\mathrm{A}_0p_{m+1}^\mathrm{B} ) \frac{(2\lambda)^2 (m+1)}{\left(\omega_{\rm A}-\omega_{\rm B}\right)^2 + (2\lambda)^2(m+1)} \sin^2{\bigg(\frac{\sqrt{\left(\omega_{\rm A}-\omega_{\rm B}\right)^2 + (2\lambda)^2(m+1)}}{2}\tau \bigg)} \\ \notag
& = \omega_{\rm A} \sum_m \frac{(2\lambda)^2(m+1)}{\left(\omega_{\rm A}-\omega_{\rm B}\right)^2 + (2\lambda)^2(m+1)}(p^\mathrm{A}_0p_{m+1}^\mathrm{B}-p^\mathrm{A}_1p_m^\mathrm{B} )
\sin^2{\bigg(\frac{\sqrt{\left(\omega_{\rm A}-\omega_{\rm B}\right)^2 + (2\lambda)^2(m+1)}}{2}\tau \bigg)}.
\end{align}
\end{widetext}

Comparing Eqs.~\eqref{eq:intEnergyBatt} and \eqref{eq:InternalEnergyActivator}, we obtain the simple relation 
\begin{align}
    \label{eq:InputOutput}
    \Delta E_{\rm A}=-\frac{\omega_{\rm A}}{\omega_{\rm B}}\Delta E_{\rm B}.
\end{align}
This relation reflects the fact that each elementary excitation transferred between the activator and battery changes their energies by different quanta, $\omega_{\rm A}$ and $\omega_{\rm B}$, respectively.

Since all remaining prefactors in Eq.~\eqref{eq:InternalEnergyActivator} are non-negative, the condition for the activator to lose energy, $\Delta E_{\rm A}\leq0$, is

\begin{align}
p^\mathrm{A}_1 p^\mathrm{B}_{m}
\geq
p^\mathrm{A}_0 p^\mathrm{B}_{m+1}.
\end{align}

Substituting the thermal occupation probabilities~\eqref{eq:populationActivator} and ~\eqref{eq:populationBattery} yields

\begin{align}
\frac{\left(
e^{-\hbar\omega_{\rm A}/T_{\rm A}}
-
e^{-\hbar\omega_{\rm B}/T_{\rm B}}
\right)
\left(
1-e^{-\hbar\omega_{\rm B}/T_{\rm B}}
\right)}
{1+e^{-\hbar\omega_{\rm A}/T_{\rm A}}}
\geq 0,
\end{align}

which simplifies to

\begin{align}
e^{-\hbar\omega_{\rm A}/T_{\rm A}}
&\geq
e^{-\hbar\omega_{\rm B}/T_{\rm B}},
\label{eq:leqEA}
\\ \notag
\frac{T_{\rm A}}{T_{\rm B}}
&\geq
\frac{\omega_{\rm A}}{\omega_{\rm B}}.
\end{align}

We now consider the regime corresponding to condition~\eqref{eq:condition2} in Sec.~\ref{sec:OscillatorQubit}, namely $T_{\rm B}<T_{\rm A}$ and $\Delta E_{\rm A}<0$. Combining this with the requirement that positive stirring work is supplied,

\begin{align}
W_{\rm stirr}
=
\Delta E_{\rm A}
+
\Delta E_{\rm B}
>0,
\end{align}

and using Eq.~\eqref{eq:InputOutput}, we obtain the additional constraint

\begin{align}
\frac{\omega_{\rm A}}{\omega_{\rm B}}
\leq 1.
\label{eq:geqEA2}
\end{align}

Conversely, the condition for the activator to gain energy, $\Delta E_{\rm A}\geq0$, is

\begin{align}
\frac{\omega_{\rm A}}{\omega_{\rm B}}
\geq
\frac{T_{\rm A}}{T_{\rm B}}.
\label{eq:geqEA3}
\end{align}

Considering condition~\eqref{eq:condition1}, namely $T_{\rm B}>T_{\rm A}$ and $\Delta E_{\rm A}>0$, together with the requirement $W_{\rm stirr}>0$, yields the additional constraint

\begin{align}
\frac{\omega_{\rm A}}{\omega_{\rm B}}
\geq 1.
\label{eq:geqEA4}
\end{align}

\nocite{*}

\bibliography{references}

\end{document}